# An Adaptive Successive Cancellation List Decoder for Polar Codes with Cyclic Redundancy Check

Bin Li, Hui Shen, and David Tse, *Fellow, IEEE*

*Abstract*—In this letter, we propose an adaptive SC (Successive Cancellation)-List decoder for polar codes with CRC. This adaptive SC-List decoder iteratively increases the list size until the decoder outputs contain at least one survival path which can pass CRC. Simulation shows that the adaptive SC-List decoder provides significant complexity reduction. We also demonstrate that polar code (2048, 1024) with 24-bit CRC decoded by our proposed adaptive SC-List decoder with very large list size can achieve a frame error rate $\text{FER} \leq 10^{-3}$ at $E_b/N_o = 1.1\text{dB}$, which is about 0.2dB from the information theoretic limit at this block length.

*Index Terms*—*Polar codes, List deocder*

## I. INTRODUCTION

Polar codes are a major breakthrough in coding theory [1]. They can achieve Shannon capacity with a simple encoder and a simple successive cancellation decoder, both with low complexity of the order of $\text{O}(N \log N)$, where $N$ is the code block size. But for short and moderate lengths, the error rate performance of polar codes with the SC decoding is not as good as LDPC or turbo codes. A new SC-list decoding algorithm was proposed for polar codes recently [2], which performs better than the simple SC decoder and performs almost the same as the optimal ML (maximum likelihood) decoding at high SNR. In order to improve the low minimum distance of polar codes, the concatenation of polar codes with simple CRC was proposed [2], and it was shown that a simple concatenation scheme of polar code (2048, 1024) with a 16-bit CRC using the SC-List decoding can outperform Turbo and LDPC codes. In this letter, we make the observation that, with the concatenation of CRC, the required list size to approach ML performance is much larger than when there is no concatenation. The complexity of the above SC-List decoder would be very high for such large list sizes. To overcome this issue, we propose an adaptive SC-List decoder whose average complexity essentially does not increase with the list size. Average complexity, as opposed to worst-case complexity, is a meaningful metric in several settings. Low average complexity means low processing energy consumption, which is important for example in a handset. In a base station where multiple users are simultaneously decoded, low average complexity to decode each user means the ability to statistically and efficiently share the processing of the users. Similarly, the processing among different decoding blocks can be shared over time.

In section II, we review both the simple SC decoder and the SC-List decoder, and in section III we propose a new adaptive SC-List decoder. Finally we draw some conclusions.

## II. SC-LIST DECODING OF POLAR CODES

### A. Polar Codes

Let $F = \begin{bmatrix} 1 & 0 \\ 1 & 1 \end{bmatrix}$, $F^{\otimes n}$ is a $N \times N$ matrix, where $N = 2^n$, $\otimes n$ denotes $n$th Kronecker power, and $F^{\otimes n} = F \otimes F^{\otimes (n-1)}$. Let the $n$-bit binary representation of integer $i$ be $b_{n-1}, b_{n-2}, ..., b_0$. The $n$-bit representation $b_0, b_1, ..., b_n$ is a bit-reversal order of $i$. The generator matrix of polar code is defined as $G_N = B_N F^{\otimes n}$, where $B_N$ is a bit-reversal permutation matrix. The polar code is generated by

$$x_1^N = u_1^N G_N = u_1^N B_N F^{\otimes n} \quad (1)$$

where $x_1^N = (x_1, x_2, ..., x_N)$ is the encoded bit sequence, and $u_1^N = (u_1, u_2, ..., u_N)$ is the encoding bit sequence. The bit indexes of $u_1^N$ are divided into two subsets: the one containing the information bits represented by $A$ and the other containing the frozen bits represented by $A^c$. The polar code can be further expressed as

$$x_1^N = u_A G_N(A) \oplus u_{A^c} G_N(A^c) \quad (2)$$

where $G_N(A)$ denotes the submatrix of $G_N$ formed by the rows with indices in $A$, and $G_N(A^c)$ denotes the submatrix of $G_N$ formed by the rows with indices in $A^c$. $u_A$ are the information bits, and $u_{A^c}$ are the frozen bits.

Bin Li and Hui Shen are with the Communications Technology Research Lab., Huawei Technologies, Shenzhen, P. R. China (e-mail:{binli.binli, hshen}@huawei.com).
David Tse is with the Dept. of Electrical Engineering and Computer Science, University of California at Berkeley, CA 94720-1170, USA (e-mail: dtse@eecs.berkeley.edu).

Polar codes can be decoded with the very efficient SC decoder, which has a decoding complexity of $O(N\log N)$ and can achieve capacity when $N$ is very large.

### B. SC-List Decoder for Polar Codes

Though the SC decoder approaches Shannon capacity, it does not perform well for polar codes with small and medium block lengths. A more powerful SC-List decoder was proposed in [2] and performs much better than the SC decoder. Instead of keeping only one survival path as in the SC decoder, the SC-List decoder keeps $L$ survival paths. At each decoding time, the SC-List decoder splits each current decoding path into two paths attempting both $\hat{u}_i = 0$ and $\hat{u}_i = 1$ (if $u_i$ is an unfrozen bit). When the number of paths grows beyond a predefined threshold $L$, the SC-List decoder discards the worst (least probable) paths, and only keeps the $L$ best paths. Simulations show that the SC-List decoder with $L=32$ performs much better than the simple SC decoder and it can achieve the same performance as the optimal ML decoder at high SNR for the polar code with coding rate $R=1/2$, and $N=2048$ and $8192$ [2].

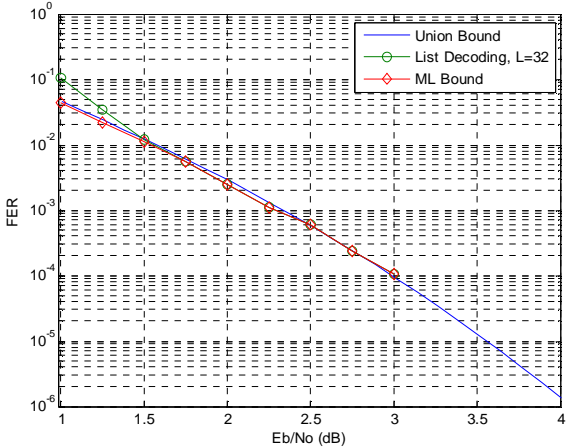

Fig. 1. The FER performance of the SC-List decoding with $L=32$ for polar code (2048, 1024).

Fig.1 shows the frame error rate (FER) of polar code (2048, 1024) using the SC-List decoding with $L=32$, where the signal is BPSK modulated and transmitted over the additive white Gaussian noise (AWGN) channel. The ML bound is also simulated in the same way as in [2]. We plot an approximate union bound using the minimum distance $d_0 = 16$ and the second least distance $d_1 = 24$, and the numbers of codewords at these distances are $N_{16} = 11648$ and $N_{24} = 215040$, respectively. We see that this approximate union bound matches the ML lower bound quite well. The parameters $d_0, d_1, N_{16}$ and $N_{24}$ are obtained as follows. We set very high $SNR \to \infty$ and use very large list sizes for the list decoder. We postulate that in this regime the list at the decoder output is most likely to contain only the least weight codewords when all zeros codeword is transmitted. Fig. 2 shows the number of codewords at each weight versus the list size. We find that there are only codewords of weight 16, 24 and 32 in the list, and we also find that $L = N_{16} + N_{24} + N_{32} + 1$ for all simulated values of $L$. To give further evidence that we found all the codewords of weight 16 and weight 24, we see in the plot that the number of the weight-16 codewords saturates at $L=11648$ and the number of the weight-24 codewords saturates at $L=215040$.

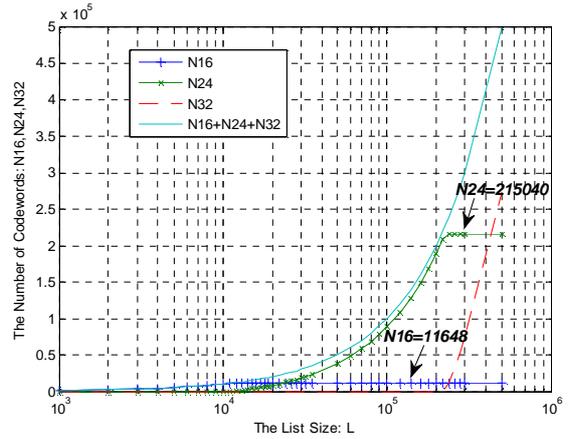

Fig. 2. The number of the least weight codewords versus the list size.

### III. A NEW ADAPTIVE SC-LIST DECODER FOR POLAR CODES WITH CRC

To improve the performance of polar codes, the concatenation of polar codes with CRC was proposed in [2]. The decoder performs SC-List decoding and then performs a CRC on each of the survival paths remaining in the list at the end of SC-List decoding. The CRC eliminates the wrong paths.

To understand why the concatenation of polar codes with CRC provides a performance gain, we did the following experiment. We took the 11648 weight-16 codewords and the 215040 weight-24 codewords we found for the polar code, and run all the 16-bit CRC's in [3] on them. We found that for five of the six CRC's (except CRC-16-DNP), ALL of the weight-16 and weight-24 codewords are eliminated by the CRC. Therefore, the 16-bit CRC's improve the minimum Hamming distance of the polar code (2048, 1024) from 16 to 32. Therefore significant improvement can be obtained. Indeed, simulations show that this concatenation of the polar code (2048, 1024) with a 16-bit CRC using the SC-List decoder performs much better than the purely non-concatenated polar codes.

Although the SC-List decoder with $L=32$ can achieve ML performance for the polar code (2048, 1024) without CRC, it turns out, for reasons to be explained later, that with CRC, a larger $L$ would be needed to exploit the increased distance benefit. Indeed, simulations show that the performance of the SC-List decoding of the polar code (2048, 1024) with a 16-bit CRC has not reached the ML bound even when $L$ is of the order of a million. Hence, in order to achieve ML performance, the polar code (2048, 1024) with a 16-bit CRC needs very large $L$, which leads to very high decoding complexity.

**Adaptive SC-List Decoder**
**Initialization:** Initialize $L=1$ for the SC-list decoder;
**Iterations:**
1  Perform the SC-List decoding, and then perform CRC on each survival path at the end of SC-List decoding.
2  If there is one or more than one paths passing CRC, output the path passing CRC and with highest probability, and exit decoding; otherwise go to 3;
3  Update $L=2\times L$. If $L\leq L_{\max}$, go to 1; otherwise output the path with the highest probability and exit decoding;

We observed that for most of the received frames, the SC-List decoder with very small $L$ can successfully decode information bits, and there are very few frames that need large $L$ for successful decoding. Therefore, in order to reduce the decoding complexity, we propose an adaptive SC-List decoder for polar codes with CRC. The adaptive SC-List decoder initially uses very small $L$, and then iteratively increases $L$ (if there is no survival path passing CRC), until $L$ reaches a predefined number $L_{\max}$.

TABLE I. THE MEAN OF $L$ OF THE ADAPTIVE SC-LIST DECODER

| $E_b/N_o$(dB) | 1.0 | 1.2 | 1.4 | 1.6 | 1.8 | 2.0 |
|---|---|---|---|---|---|---|
| $L_{\max}$=32 | 16.64 | 8.03 | 3.86 | 2.04 | 1.39 | 1.14 |
| $L_{\max}$=128 | 35.31 | 12.16 | 4.52 | 2.17 | 1.41 | |
| $L_{\max}$=512 | 70.41 | 19.14 | 5.45 | 2.27 | | |
| $L_{\max}$=2048 | 133.40 | 30.80 | 6.64 | 2.36 | | |
| $L_{\max}$=8192 | 271.07 | 52.59 | 7.88 | 2.47 | | |

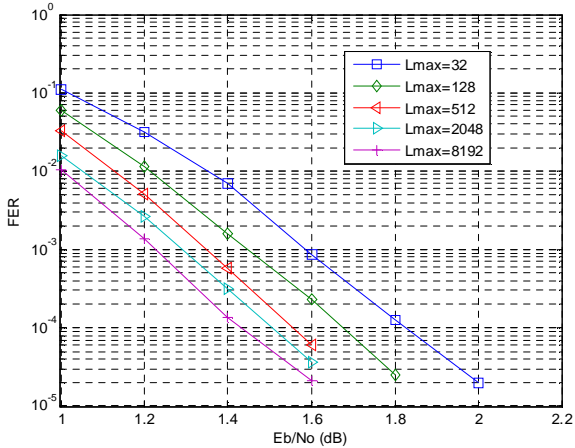

Fig. 3. The FER performance of the polar code (2048, 1024) with 16-bit CRC using the adaptive SIC-List decoder with different $L_{\max}$.

Fig.3 shows the FER performance of the adaptive SC-List decoder for the polar code (2048, 1024) with a 16-bit CRC. The channel is an AWGN channel, and the signal is modulated by BPSK modulation. It is shown that there is about 0.4dB gain of $L_{\max}=8192$ over $L_{\max}=32$ at $FER=10^{-3}$. Since the list most likely contains the weight-16 and weight-24 codewords, and these codewords cannot pass CRC. The frame error rate is in fact dominated by the probability that the correct path is not in the list. When we increase $L$, we essentially increase the probability that the correct path is in the list, and therefore we can get better performance. We predict that, in order to achieve ML decoding, $L>N_{16}+N_{24}+1$.

When the adaptive list decoder with $L<L_{\max}$ contains the correct path, then the non-adaptive list decoder with $L=L_{\max}$ will also contain the correct path, and both decoders can successfully decode; If the adaptive list decoder with $L<L_{\max}$ does not contain the correct path, it will increase $L$ until $L=L_{\max}$, this leads to that both decoders use the same $L=L_{\max}$ and the two decoders perform the same. Therefore the performance of the adaptive SC-List decoder with $L_{\max}=32$ and $L_{\max}=8192$ are the same as the non-adaptive SC-List decoder with constant $L=32$ and $L=8192$, respectively.

TABLE I shows the mean $L$ for different $E_b/N_o$ and different $L_{\max}$. With the increasing of $E_b/N_o$, the SC-List decoder is more likely to successfully decode the received frames with the same $L$, and therefore the mean of $L$ becomes smaller for the adaptive SC-List decoder. Since the complexity of the SC-List decoder is linear in the list size, the SC-List decoder with constant $L$ has a complexity of the order of $O(LN\log N)$ and our adaptive SC-List decoder has an average complexity of the order of $O(\overline{L}N\log N)$. It is seen that under $L_{\max}=32$, the mean of $L$ is $\overline{L}=2.04$ for $E_b/N_o=1.6\text{dB}$; this is about 16 times complexity reduction but with the same performance compared with the constant $L=32$. The mean of $L$ under $L_{\max}=8192$ is $\overline{L}=2.47$ for $E_b/N_o=1.6\text{dB}$; this is about 3316 times complexity reduction but with the same performance compared with the constant $L=8192$.

It is interesting to mention that we simulated the adaptive SC-List decoder with very large $L_{\max}=262144$ for the polar code (2048, 1024) with a 24-bit CRC, the mean $\overline{L}=818.5$. We found that this concatenated code can achieve $FER\leq 10^{-3}$ at $E_b/N_o=1.1\text{dB}$. To compare this performance to the Shannon limit at the same block length, we use a result from [4]. It says that for a wide range of channels, the maximum rate that can be achieved at block length $N$ and $FER(e)$ can be well-approximated by the formula:

$$R = C - \frac{V}{\sqrt{N}} Q^{-1}(e) \quad (3)$$

where $C$ is the channel capacity and $V$ is a quantity called the channel dispersion that can be computed from the channel statistics. Applying this result to the binary input AWGN channel, we calculated that to achieve a rate $R = 0.5$ and $\text{FER} = 10^{-3}$, the minimum $E_b/N_o$ required is 0.9dB. Hence our system is only 0.2 dB from the Shannon limit. This performance is much more difficult to be simulated by the SC-List decoder with constant $L = 262144$.

## IV. CONCLUSION

In this letter, we propose an adaptive SC-List decoder which uses adaptive list size $L$ instead of constant $L$ for the decoding process. Compared with the conventional SC-List decoder with constant $L$, the adaptive SC-List decoder can achieve the same performance but with significantly lower complexity; alternatively it can achieve much better performance but with the same decoding complexity. Using this adaptive SC-List decoder with very large $L_{\max} = 262144$ (with $\bar{L} = 818.5$) for the polar code (2048, 1024) with a 24-bit CRC, we can achieve $\text{FER} \leq 10^{-3}$ at $E_b/N_o = 1.1\text{dB}$, which is about 0.2 dB from the information theoretic limit at the same block length.

## REFERENCES

[1] E. Arıkan, "Channel polarization: A method for constructing capacity achieving codes for symmetric binary-input memoryless channels," IEEE Trans. Inform. Theory, vol. 55, pp. 3051–3073, July 2009.
[2] I. Tal and A. Vardy, "List Decoding of Polar Codes," available as online as arXiv: 1206.0050v1.
[3] Wikipedia, Cyclic redundancy check, http://en.wikipedia.org/wiki/Cyclic_redundancy_check (accessed on Aug. 15, 2012.)
**[4]** Y. Polyanskiy, H. Vincent Poor, and S. Verdú, "Channel Coding Rate in the Finite Blocklength Regime", IEEE Trans. Inform. Theory, vol. 56, No. 5, May 2010.
[5] E. Arıkan, "Polar Coding: Status and Prospects", Plenary Talk of IEEE International Symposium on Inform. Theory, Saint Petersburg, Russia, 2011.
[6] S. B. Korada, E. Sasoglu, and R. L. Urbanke, "Polar codes: Characterization of exponent, bounds, and constructions," IEEE Trans. Inform. Theory, vol. 56, pp. 6253–6264, 2010.
[7] E. Arıkan and E. Telatar, "On the rate of channel polarization," in Proc.IEEE Int'l Symp. Inform. Theory , Seoul, South Korea, 2009, pp. 1493–1495.
[8] I. Tal and A. Vardy, "How to construct polar codes," in Proc. IEEE Inform. Theory Workshop, Dublin, Ireland, 2010.